\documentclass[%
 reprint,
superscriptaddress,
amsmath,amssymb,
aps,
pra,
]{revtex4-1}
\usepackage{caption}
\usepackage{subcaption}
\usepackage{graphicx}
\usepackage{xcolor}
\usepackage{graphicx}
\usepackage{dcolumn}
\usepackage{bm}

\begin{document}

\preprint{APS/123-QED}

\title{Refraction of space-time wave packets in a dispersive medium}

\author{Murat Yessenov}
\author{Sanaz Faryadras}
\author{Sepehr Benis}
\author{David J. Hagan}
\author{Eric W. Van Stryland}
\author{Ayman F. Abouraddy}
\thanks{corresponding author: raddy@creol.ucf.edu}

\affiliation{CREOL, The College of Optics \& Photonics, University of Central Florida, Orlando, FL 32816, USA}




\begin{abstract}
Space-time (ST) wave packets are a class of pulsed optical beams whose spatio-temporal spectral structure results in propagation invariance, tunable group velocity, and fascinating refractive phenomena. Here, we investigate the refraction of ST wave packets at a planar interface between two dispersive, homogeneous, isotropic media. We formulate a new refractive invariant for ST wave packets in this configuration, from which we obtain a law of refraction that determines the change in their group velocity across the interface. We verify this new refraction law in ZnSe and CdSe, both of which manifest large chromatic dispersion at near-infrared frequencies in the vicinity of their band edges. ST wave packets can thus be a tool in nonlinear optics for bridging large group-velocity mismatches in highly dispersive scenarios.
\end{abstract}



\maketitle

Introducing precise spatio-temporal spectral structure into a pulsed beam can profoundly modify its propagation characteristics \cite{Turunen10PO,FigueroaBook14}. For example, space-time (ST) wave packets \cite{Yessenov19OPN} exhibit unique characteristics by virtue of their spatio-temporal spectral structure \cite{Donnelly93ProcRSLA,Saari04PRE,Kondakci16OE,Parker16OE,Porras17OL,Efremidis17OL,Wong17ACSP2,Kondakci19OL}, including propagation-invariance \cite{Valtna07OC,Zamboni09PRA,Kondakci17NP,Kondakci18PRL,Bhaduri19OL}, tunable group velocities \cite{Salo01JOA,Kondakci19NC}, axial acceleration \cite{Yessenov2020PRLaccel}, ST Talbot self-imaging \cite{Hall21APLSTTalbot}, and self-healing \cite{Kondakci18OL}. In contrast to earlier proposed localized waves \cite{Turunen10PO,FigueroaBook14}, the group velocity of ST wave packets can depart significantly from that of conventional pulsed beams while remaining in the paraxial regime \cite{Yessenov19PRA}. This broad tunability is possible because only low spatial frequencies are required in the synthesis process, and we thus refer to these ST wave packets as `baseband'. Until recently, only `sideband' wave packets such as Brittingham's focus-wave mode \cite{Brittingham83JAP} and X-waves \cite{Saari97PRL} were accessible, in which the low spatial frequencies are excluded on physical grounds \cite{Yessenov19PRA}, and whose reported group velocities consequently deviate only minutely from $c$ (the speed of light in vacuum) \cite{Bowlan09OL}.

\begin{figure}[t!]
\centering
\includegraphics[width=8.6cm]{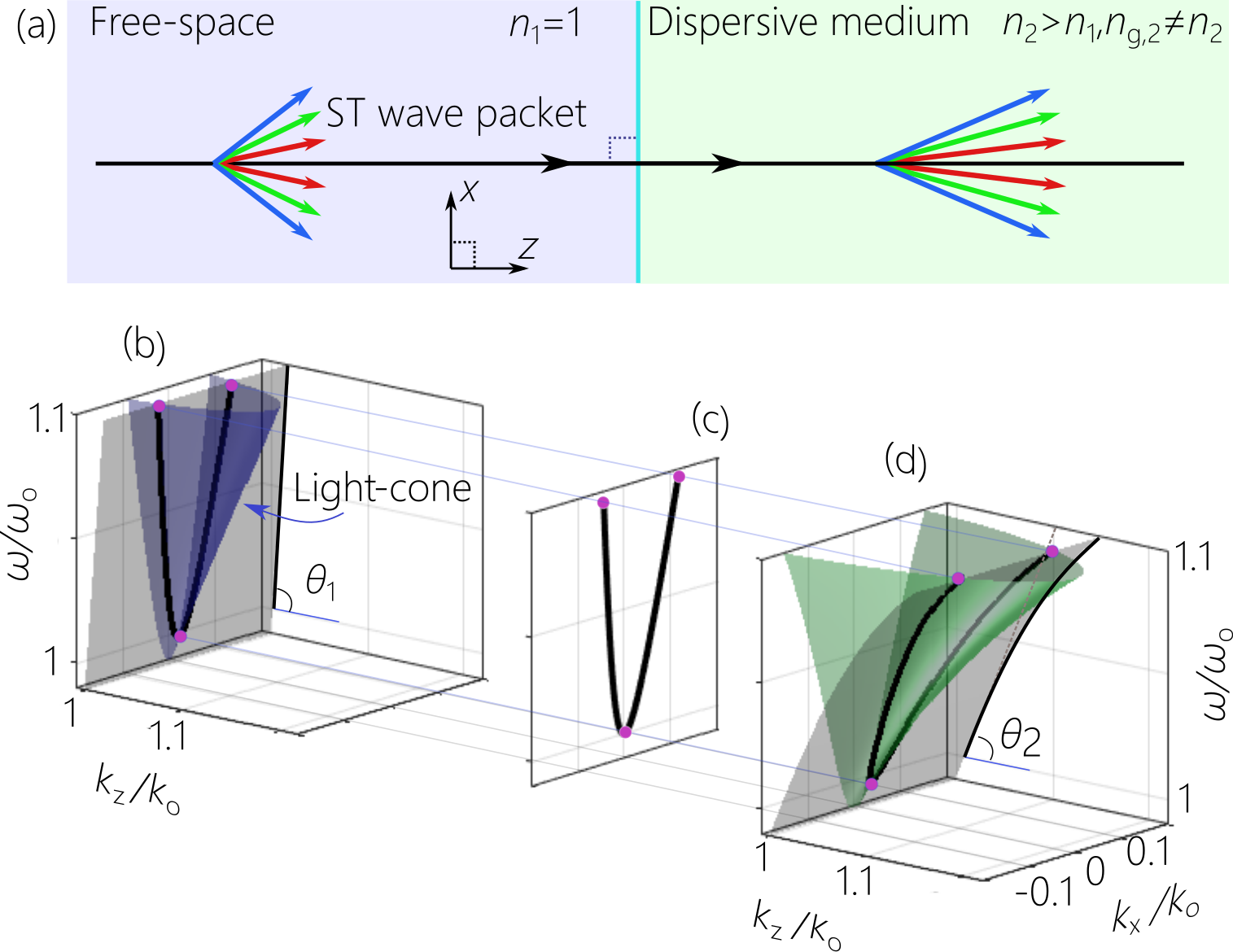}
\caption{(a) Refraction of ST wave packets at normal incidence on a planar interface between free space and a dispersive medium. (b) The spectral support domain of a ST wave packet in free space is the intersection of the light-cone $k_{x}^{2}+k_{z}^{2}\!=\!(\tfrac{\omega}{c})^{2}$ with a plane. The spectral projection onto the $(k_{z},\tfrac{\omega}{c})$-plane is a straight line making an angle $\theta_{1}$ with the $k_{z}$-axis. (c) The spectral projection from (b) onto the $(k_{x},\tfrac{\omega}{c})$-plane is invariant across the interface. (d) In the dispersive medium $k_{x}^{2}+k_{z}^{2}\!=\!n^{2}(\omega)(\tfrac{\omega}{c})^{2}$, the spectral projection onto the $(k_{z},\tfrac{\omega}{c})$-plane makes an angle $\theta_{2}$ with the $k_{z}$-axis.}
\label{Fig:Concpet}
\end{figure}

Previous theoretical studies have shown that propagation invariance can be maintained for baseband ST wave packets in dispersive media \cite{Longhi04OL,Porras04OL,Malaguti08OL,Malaguti09PRA}. However, because such wave packets could not be synthesized, those results have informed experiments that \textit{produce} ST wave packets via nonlinear interactions \cite{DiTrapani03PRL}, rather than coupling ST wave packets synthesized in free space to the medium [Fig.~\ref{Fig:Concpet}(a)]. This hurdle has now been cleared, and the recently developed spatio-temporal Fourier synthesis methodology can readily produce baseband ST wave packets \cite{Kondakci17NP,Yessenov19OPN}. Nevertheless, careful consideration must be paid to refraction at the medium interface. Indeed, recent studies of the refraction of baseband ST wave packets at a planar interface between \textit{non-dispersive} media have revealed fascinating refractive phenomena, such as group-velocity invariance, anomalous refraction, group-velocity inversion \cite{Bhaduri20NP,Yessenov21JOSAA1,Motz21JOSAA2}, incident-angle-dependent group velocity of the transmitted wave packet \cite{Yessenov21JOSAA3}, blind synchronization \cite{Bhaduri20NP,Yessenov21JOSAA3}, and isochronous ST wave packets \cite{Motz21OL} -- all of which are absent from focus-wave modes \cite{Hillion98JO} and X-waves \cite{Salem12JOSAA}. Undergirding these phenomena is a refractive invariant associated with ST wave packets -- the `spectral curvature' -- that leads to a law of refraction governing the change in their group velocity across an interface. This invariant vanishes in the case of conventional pulsed fields, thus precluding the observation of these refractive phenomena.

Here, we study the refraction of ST wave packets normally incident at a planar interface between two dispersive media [Fig.~\ref{Fig:Concpet}(a)]. We find that dispersion modifies the refractive invariant for ST wave packets and thus modifies the law of refraction for ST wave packets in the presence of chromatic dispersion. Whereas the spectral curvature in the absence of dispersion depends on the refractive index at the operating wavelength, the new spectral curvature in the presence of dispersion depends on both the refractive index \textit{and} the group index, both evaluated at the operating wavelength. We verify this law of refraction with ST wave packets at a wavelength of 775~nm incident from free space onto ZnSe (refractive index $n\!\approx\!2.5$ and group index $n_{\mathrm{g}}\!\approx\!2.7$) and onto CdSe ($n\!\approx\!2.6,n_{\mathrm{g}}\!\approx\!3.7$), which are highly dispersive at this wavelength. Such an approach allows for unprecedented tunability of the group velocity for a ST wave packet in a dispersive medium, which can help reduce the axial walk-off due to group-velocity mismatch between disparate wavelengths close to a band edge where dispersion is high.

We start by describing ST wave packets and their refraction in a dispersion-free medium, whereupon the relationship $k_{x}^{2}+k_{z}^{2}\!=\!n_{\mathrm{o}}^{2}(\tfrac{\omega}{c})^{2}$holds, which is represented geometrically by the surface of a `light-cone' [Fig.~\ref{Fig:Concpet}(b)]; here $n_{\mathrm{o}}$ is the refractive index, $k_{x}$ and $k_{z}$ are the transverse and axial components of the wave vector along $x$ and $z$, respectively, $\omega$ is the temporal frequency, and we take the field to be uniform along $y$. The spectral support domain for a propagation-invariant ST wave packet is the intersection of the light-cone with a plane $\Omega\!=\!(k_{z}-n_{\mathrm{o}}k_{\mathrm{o}})\widetilde{v}$, which is parallel to the $k_{x}$-axis, and makes an angle $\theta$ with the $k_{z}$-axis; here $\Omega\!=\!\omega-\omega_{\mathrm{o}}$, $\omega_{\mathrm{o}}$ is a fixed carrier frequency, and $k_{\mathrm{o}}\!=\!\omega_{\mathrm{o}}/c$ is the associated wave number. This construction yields a propagation-invariant ST wave packet that travels rigidly in the medium at a group velocity $\widetilde{v}\!=\!c\tan{\theta}\!=\!c/\widetilde{n}$ determined by the spectral tilt angle $\theta$, where $\widetilde{n}\!=\!\cot{\theta}$ is the group index \cite{Yessenov19PRA,Yessenov19OE}. In principle, the group velocity can take on arbitrary values: subluminal ($\widetilde{v}\!<\!c/n$ and $\widetilde{n}\!>\!n$) or superluminal ($\widetilde{v}\!>\!c/n$ and $\widetilde{n}\!<\!n$). The spectral projection onto the $(k_{z},\tfrac{\omega}{c})$-plane is a straight line [Fig.~\ref{Fig:Concpet}(b)], thus indicating dispersion-free propagation, and the projection onto the $(k_{x},\tfrac{\omega}{c})$-plane can be approximated by a parabola in the vicinity of $k_{x}\!=\!0$ [Fig.~\ref{Fig:Concpet}(c)].

This constraint in the narrowband ($\Delta\omega\!\ll\!\omega_{\mathrm{o}}$) paraxial ($\Delta k_{x}\!\ll\!k_{\mathrm{o}}$) regime, $k_{z}\!\approx\!n_{\mathrm{o}}k_{\mathrm{o}}+n_{\mathrm{o}}\tfrac{\Omega}{c}-\tfrac{k_{x}^{2}}{2k_{\mathrm{o}}}$, yields:
\begin{equation}\label{Eq:STcorrelationNondispersive}
\tfrac{1}{2}(k_{x}/k_{\mathrm{o}})^{2}(\omega_{\mathrm{o}}/\Omega)\approx n_{\mathrm{o}}(n_{\mathrm{o}}-\widetilde{n}).
\end{equation}
Because $k_{x}$ and $\Omega$ are invariant across a planar interface [Fig.~\ref{Fig:Concpet}(c)], the quantity $n_{\mathrm{o}}(n_{\mathrm{o}}-\widetilde{n})$, which is related to the curvature of the parabolic spectral projection onto the $(k_{x},\tfrac{\omega}{c})$-plane, is a \textit{refractive invariant}. In other words, $n_{\mathrm{o}}(n_{\mathrm{o}}-\widetilde{n})$ for a ST wave packet is in fact independent of the medium, and when a ST wave packet traverses an interface between two media of refractive indices $n_{1}$ and $n_{2}$, the group velocity changes (from $\widetilde{v}_{1}\!=\!c/\widetilde{n}_{1}$ to $\widetilde{v}_{2}\!=\!c/\widetilde{n}_{2}$ in the second medium) to maintain the spectral curvature fixed \cite{Bhaduri20NP}:
\begin{equation}\label{Eq:LoRinNonDispersive}
n_{1}(n_{1}-\widetilde{n}_{1})=n_{2}(n_{2}-\widetilde{n}_{2}).
\end{equation}
In contrast to conventional wave packets whose group velocity depends solely on the local optical properties of the medium, the refraction of ST wave packets features a `memory effect': the group velocity of the transmitted ST wave packet depends also on the group velocity of the incident wave packet, and the refractive indices of both media \cite{Bhaduri19Optica,Bhaduri20NP}. This unique memory effect has fascinating consequences, including: (1) group-velocity invariance, whereby the ST wave packet retains its group velocity $\widetilde{n}_{2}\!=\!\widetilde{n}_{1}$ whenever $\widetilde{n}_{1}\!=\!n_{1}+n_{2}$, regardless of the index contrast; (2) anomalous refraction, whereby the group-velocity increases in a higher-index media whenever $\widetilde{n}_{1}\!>\!n_{1}+n_{2}$; and (3) group-velocity inversion $\widetilde{n}_{2}\!=\!-\widetilde{n}_{1}$ whereby the transmitted ST wave packet maintains the magnitude of the incident group velocity but switches its sign whenever $\widetilde{n}_{1}\!=\!n_{1}-n_{2}$ \cite{Bhaduri20NP,Yessenov21JOSAA1,Motz21JOSAA2}. Furthermore, at oblique incidence, $\widetilde{v}_{2}$ also depends on the incident angle \cite{Bhaduri20NP,Yessenov21JOSAA3}.

\begin{figure}[t!]
\centering
\includegraphics[width=8.6cm]{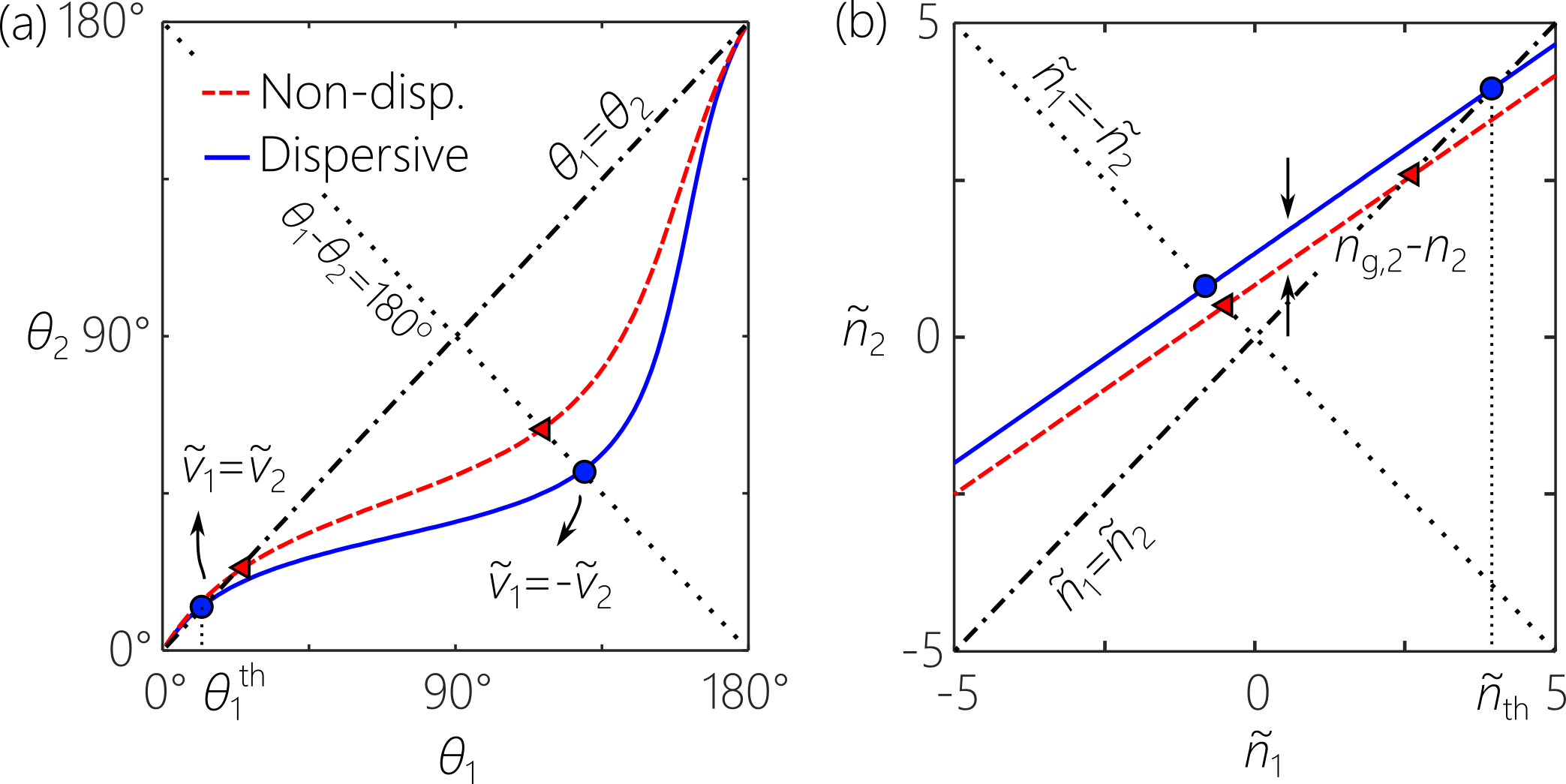}
\caption{The law of refraction for ST wave packets incident from free space ($n_{1}\!=\!n_{\mathrm{g},1}=\!1$) onto dispersive ($n_{2}\!=\!1.5$ and $n_{\mathrm{g},2}=\!2$; solid curve) and non-dispersive media ($n_{2}\!=\!n_{\mathrm{g},2}\!=\!1.5$; dashed curve), plotted in terms of (a) the spectral tilt angles $\theta_{1}$ and $\theta_{2}$, and (b) the group indices $\widetilde{n}_{1}$ and $\widetilde{n}_{2}$ of the incident and transmitted wave packets, respectively. The intersection with the diagonal $\theta_{1}\!=\!\theta_{2}$ corresponds to group-velocity invariance $\widetilde{v}_{1}\!=\!\widetilde{v}_{2}$, and the intersection with the anti-diagonal $\theta_{1}+\theta_{2}\!=\!180^{\circ}$ corresponds to group-velocity inversion $\widetilde{v}_{1}\!=\!-\widetilde{v}_{2}$.}
\label{Fig:Theory}
\end{figure}

In the presence of dispersion $n\!=\!n(\omega)$, Eq.~\ref{Eq:LoRinNonDispersive} must be modified. Here $k_{x}^{2}+k_{z}^{2}\!=\!k^{2}$ [Fig.~\ref{Fig:Concpet}(d)], and $k\!=\!n(\omega)\omega/c$ is expanded to first order $k\!\approx\!n_{\mathrm{o}}k_{\mathrm{o}}+\Omega/v_{\mathrm{g}}$, where $v_{\mathrm{g}}\!=\!c/n_{\mathrm{g}}$ is the group velocity in the medium, $n_{\mathrm{g}}$ is the medium group index evaluated at $\omega_{\mathrm{o}}$, and we neglect higher-order dispersion terms. Combining the narrowband paraxial approximation with the constraint $k_{z}\!=\!n_{\mathrm{o}}k_{\mathrm{o}}+\tfrac{\Omega}{\widetilde{v}}$ for the ST wave packet yields:
\begin{equation}\label{Eq:STcorrelationDispersive}
\tfrac{1}{2}(k_{x}/k_{\mathrm{o}})^{2}(\omega_{\mathrm{o}}/\Omega)\approx n_{\mathrm{o}}(n_{\mathrm{g}}-\widetilde{n}),
\end{equation}
and the quantity $n_{\mathrm{o}}(n_{\mathrm{g}}-\widetilde{n})$ is the new refractive invariant, which depends on 3 quantities: the medium refractive index $n_{\mathrm{o}}$, the group index $n_{\mathrm{g}}$ (both evaluated at $\omega_{\mathrm{o}}$), and the group index $\widetilde{n}$ of the ST wave packet. The invariant spectral curvature in Eq.~\ref{Eq:STcorrelationDispersive} is the product of two terms: (1) $n_{\mathrm{g}}-\widetilde{n}$ that quantifies the deviation of the group velocity of the ST wave packet from the native group velocity of the medium at the same wavelength; and (2) $n_{\mathrm{o}}$ determines the curvature of the light cone at $\omega_{\mathrm{o}}$. Of course, in the absence of chromatic dispersion, $n_{\mathrm{g}}\!\rightarrow\!n_{\mathrm{o}}$ we retrieve the invariant in Eq.~\ref{Eq:STcorrelationNondispersive}, and thus the law of refraction in Eq.~\ref{Eq:LoRinNonDispersive}. 

Utilizing the refractive invariant in Eq.~\ref{Eq:STcorrelationDispersive}, we formulate a law of refraction for ST wave packets in dispersive media:
\begin{equation}\label{Eq:LoRinDispersive}
n_{1}(n_{\mathrm{g},1}-\widetilde{n}_{1})=n_{2}(n_{\mathrm{g},2}-\widetilde{n}_{2}).
\end{equation}
We compare in Fig.~\ref{Fig:Theory} the law of refraction for ST wave packets traversing a planar interface from free space to a dispersive medium (Eq.~\ref{Eq:LoRinDispersive}), and to a non-dispersive medium (Eq.~\ref{Eq:LoRinNonDispersive}) using the same refractive indices $n_{1}$ and $n_{2}$ to highlight their distinction. The refractive phenomena displayed by ST wave packets in non-dispersive media are preserved in their dispersive counterparts after appropriate modifications: group-velocity invariance $\widetilde{n}_{1}\!=\!\widetilde{n}_{2}$ occurs when $\widetilde{n}_{1}\!=\!\tfrac{n_{\mathrm{g},1}n_{1}-n_{\mathrm{g},2}n_{2}}{n_{1}-n_{2}}$, and group-velocity inversion $\widetilde{n}_{1}\!=\!\!-\widetilde{n}_{2}$ occurs when $\widetilde{n}_{1}\!=\!\frac{n_{g,1}n_{1}-n_{g,2}n_{2}}{n_{1}+n_{2}}$. 

To verify the law of refraction in Eq.~\ref{Eq:LoRinDispersive}, we synthesize ST wave packets in free space  ($n_{1}\!=\!n_{\mathrm{g},1}\!=\!1$) and tune their group velocity before directing them to planar samples of ZnSe and CdSe. The ZnSe sample is a 5-mm-thick slab with index $n_{2}\!=2.5$ and group index $\!n_{\mathrm{g},2}\!=\!2.7$, respectively, at $\lambda_{\mathrm{o}}\!\approx\!775$~nm \cite{Marple64JoAP}. The CdSe sample is 3-mm thick with $n_{2}\!=\!2.6$ and $\!n_{\mathrm{g},2}\!=\!3.7$, as determined by spectroscopic ellipsometry (Woollam M-2000). These two samples have a large gap between $n_{2}$ and $n_{\mathrm{g},2}$ because $\lambda_{\mathrm{o}}\approx\!775$~nm is close to their corresponding band edges.  

\begin{figure}[t!]
\centering
\includegraphics[width=8.6cm]{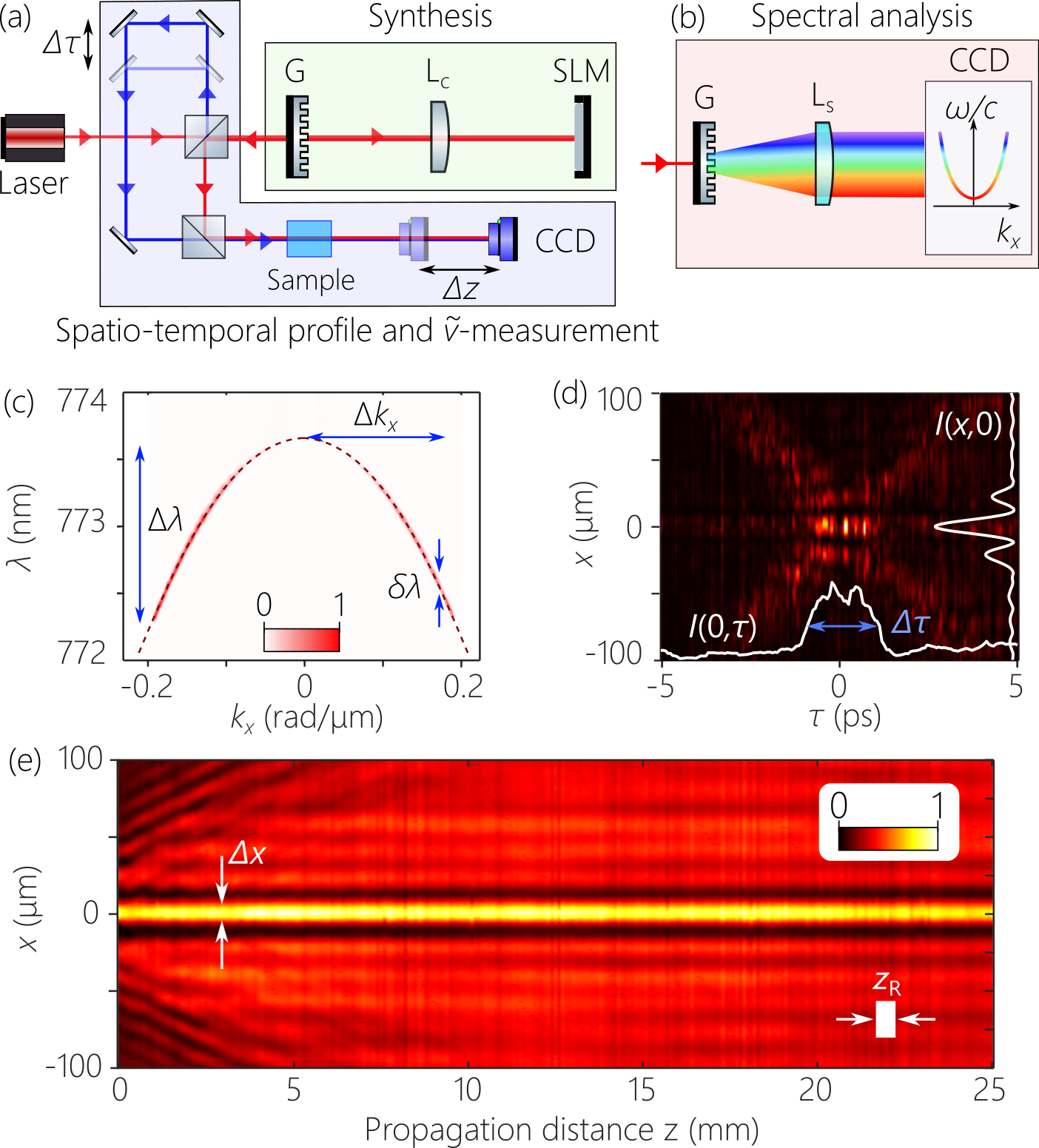}
\caption{(a) Experimental setup for the synthesis and characterization of ST wave packets. G: Diffraction grating, L$_{\mathrm{c}}$: cylindrical lens, L$_{\mathrm{s}}$: spherical lens. (b) Setup to acquire the spectra projection onto the $(k_{x},\lambda)$-plane, and (c) the measured spatio-temporal spectrum for a superluminal ST wave packet with $\theta_{1}=50^{\circ}$, $\Delta\lambda\!\approx\!1$~nm, $\Delta k_{x}\!\approx\!0.2$~rad/$\mu$m, and spectral uncertainty $\delta\lambda\!\sim\!25$~pm. (d) Measured spatio-temporal intensity profile $I(x,\tau)$ of the ST wave packet from (c) at $z\!=\!0$, the on-axis pulse-width is $\Delta\tau\!\approx\!2$~ps, and the white curves are $I(0,\tau)$ and $I(x,0)$. (e) Time-averaged intensity $I(x,z)$ along $z$ for the ST wave packet from (c). The beam size is $\Delta x\!=\!11~\mu$m, and the white bar is the Rayleigh range of a Gaussian beam with the same spatial width $\Delta x$.}
\label{Fig:Setup}
\end{figure}

The setup to synthesize ST wave packets is shown in Fig.~\ref{Fig:Setup}(a), which is based on our recent work \cite{Kondakci19NC}. Starting with plane-wave femtosecond pulses from a mode-locked laser (Clark-MXR, CPA 2010; pulses of width $\approx\!150$~fs and bandwidth $\Delta\lambda\!\approx\!7$~nm centered at $\lambda_{\mathrm{o}}\approx\!775$~nm), we spatially resolve the spectrum via a diffraction grating (1200~lines/mm) and a cylindrical lens ($f\!=\!40$~cm). At the focal plane of the lens, a reflective, phase-only spatial light modulator (SLM; Meadowlark $1920\!\times\!1152$ series) assigns a prescribed pair of spatial frequencies $\pm k_{x}(\lambda)$ to each wavelength $\lambda$ in accordance with Eq.~\ref{Eq:STcorrelationDispersive}. The retro-reflected field is reconstituted at the grating and the ST wave packet formed. Modifying the SLM phase pattern allows tuning the spectral tilt angle $\theta_{1}$, and hence the group index $\widetilde{n}_{1}=\cot{\theta_{1}}$ of the ST wave packet. The spatio-temporal spectrum projected onto the $(k_{x},\lambda)$-plane is acquired via a spatio-temporal Fourier transform [Fig.~\ref{Fig:Setup}(b)]. The measured spectrum is plotted in Fig.~\ref{Fig:Setup}(c) for a superluminal ST wave packet with spectral tilt angle $\theta_{\mathrm{1}}\!=\!50^{\circ}$ ($\widetilde{v}_{1}\!\approx\!1.19c$, $\widetilde{n}_{1}\!\approx\!0.84$).

The intensity profile $I(x,\tau)$ at a fixed axial plane $z$ and the group velocity $\widetilde{v}$ are obtained by placing the ST synthesis setup in one arm of a Mach-Zehnder interferometer, and using the original laser pulse as a reference in the other arm containing an optical delay line $\tau$ \cite{Kondakci19NC}; see Fig.~\ref{Fig:Setup}(a). We reconstruct $I(x,\tau)$ [Fig.~\ref{Fig:Setup}(d)] from the visibility of the spatially resolved interference fringes resulting from overlapping the ST wave packet (pulsewidth $\Delta\tau\!\sim\!2$~ps) with the reference pulse ($\sim\!150$~fs) at a CCD camera while sweeping $\tau$. Because the ST wave packet travels in free space at a group velocity $\widetilde{v}_{1}$, whereas the reference pulses travel at $c$, their differential group delay results in axial walk-off after displacing the CCD along $z$, which eliminates the interference. However, the interference is restored after inserting a delay $\tau$ in the reference arm, from which we estimate $\widetilde{v}_{1}$. The group velocity of the ST wave packet in the medium is estimated by first overlapping the two wave packets at the CCD and then placing the sample in their common path [Fig.~\ref{Fig:Setup}(a)]. The ST wave packet travels in the medium at $\widetilde{v}_{2}\!=\!c/\widetilde{n}_{2}$ and the reference pulses at $v_{\mathrm{g},2}\!=\!c/v_{\mathrm{g},2}$. The group-delay difference eliminates the interference of the two wave packets until it is compensated by a delay $\tau$ in the reference arm, from which we estimate $\widetilde{v}_{2}$ and $\widetilde{n}_{2}$. Finally, we confirm the diffraction-free propagation of the ST wave packets by recording the axial evolution of their intensity profile $I(x,z)$ [Fig.~\ref{Fig:Setup}(e)].

\begin{figure}[t!]
\centering
\includegraphics[width=8.6cm]{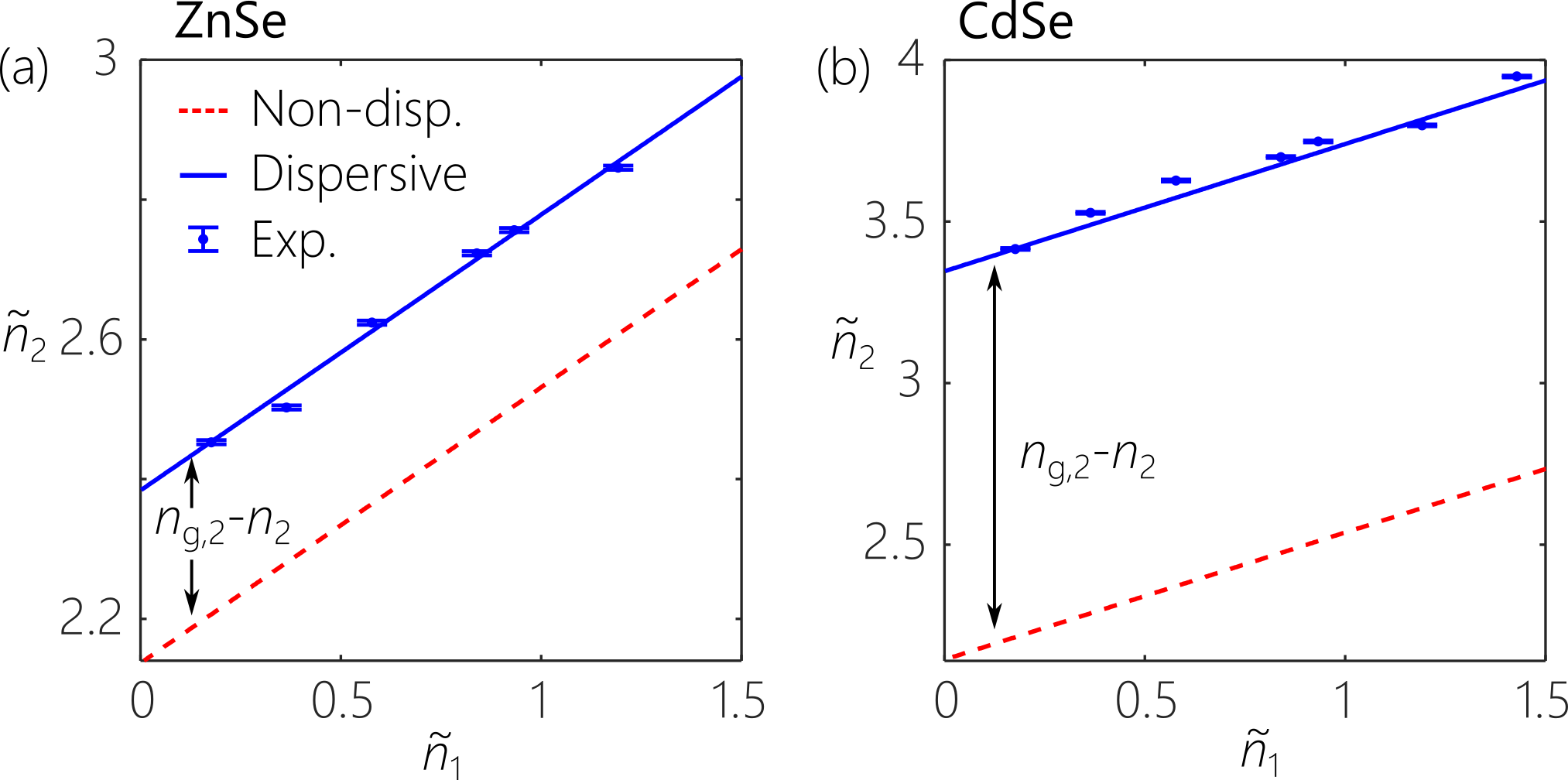}
\caption{ Experimental verification of the law of refraction for ST wave packets in a dispersive medium. (a) Measured $\widetilde{n}_{2}$ while tuning $\widetilde{n}_{1}$ for a ST wave packet incident from free space onto ZnSe, and (b) onto CdSe. The points are data, the blue solid curve is the prediction based on Eq.~\ref{Eq:LoRinDispersive}, and the red dashed curve based on Eq.~\ref{Eq:LoRinNonDispersive}. The error-bars represent the measurement uncertainty, which is dominated by the pulse-width of the ST wave packet.}
\label{Fig:Data}
\end{figure}

In Fig.~\ref{Fig:Data} we plot the measured group indices for the ST wave packets in ZnSe [Fig.~\ref{Fig:Data}(a)] and in CdSe [Fig.~\ref{Fig:Data}(b)], for normal incidence from free space. The measurements follow closely the theoretical expectations from Eq.~\ref{Eq:LoRinDispersive}. For comparison, we add the prediction based on the law of refraction for non-dispersive media (Eq.~\ref{Eq:LoRinNonDispersive}), which is displaced from the data by $\Delta\widetilde{n}_{2}\!=\!n_{\mathrm{g},2}-n_{2}\!\approx\!0.25$ in the case of ZnSe, and by $\Delta\widetilde{n}_{2}\!\approx\!1.15$ for CdSe, as expected from Eq.~\ref{Eq:LoRinNonDispersive} and Eq.~\ref{Eq:LoRinDispersive}. In both dispersive media, the transmitted light travels slower that the incident light ($\widetilde{n}_{2}>\widetilde{n}_{1}$, $\widetilde{v}_{2}<\widetilde{v}_{1}$) over the whole range of measurements, as expected in the normal refraction regime.

The wide tunability of the group velocity of a ST wave packet independently of the medium group index is particularly useful for nonlinear interactions involving disparate wavelengths. A particularly interesting example is non-degenerate two-photon absorption, which can lead to the efficient absorption of a probe photon near the band edge when accompanied by a far-from-resonance photon. However, in most materials the difference in the group indices for such non-degenerate photons is extremely large. For example, in CdSe (bandgap $E_{\mathrm{g}}\!=\!1.7$~eV) the group index at $\lambda_{\mathrm{o}}\!=\!775$~nm close to the band edge is $n_{\mathrm{g}}\!\approx\!3.7$, whereas that at $\lambda\!=\!1900$~nm is $n_{\mathrm{g}}\!\approx\!2.6$. This large difference in group indices reduces the interaction length to $\approx\!250$~$\mu$m for $\sim\!1$-ps pulses \cite{Negres02JQE}. It is challenging to compensate for such a large mismatch in group velocities along the propagation axis using traditional approaches (tilted pulse fronts require a noncollinear configuration \cite{Torres10AOP}). ST wave packets can be potentially utilized to address this challenge by making use of a ST wave packet as the probe. One can then change the group index of the probe (at $\lambda_{\mathrm{o}}\!=\!775$~nm) by $\Delta n_{\mathrm{g}}\!\approx\!1.1$ to match the group index of a conventional pump pulse at $\lambda\!=\!1900$~nm.

In conclusion, we have formulated theoretically and verified experimentally a law of refraction for ST wave packets in dispersive media. In the presence of chromatic dispersion, the group velocity of the transmitted wave packet depends on the refractive \textit{and} group indices of both media, as well as the group velocity of the incident ST wave packets. The unique refractive phenomena displayed by ST wave packets in non-dispersive media are preserved in the presence of dispersion. We verified this law of refraction in highly dispersive ZnSe and CdSe, both of which have a large gap between their refractive index and group index at the operating wavelength. Results reported in this work are crucial for applications of ST wave packets in nonlinear and quantum optics; for example, group-velocity matching of pump and probe pulses in non-degenerate two-photon absorption \cite{Fishman11NP}. Finally, we have examined the impact of the first-order dispersion term (the group index), but not the second-order term (group-velocity dispersion), which we will pursue elsewhere. In particular, the dynamics exhibited by ST wave packets while changing their group velocity in dispersive media as predicted in  \cite{Porras03PRE,Porras05JOSAB,Malaguti08OL,Malaguti09PRA} can now be put to test.

\textit{Note}. After completing our experiments, a new theoretical study \cite{He21Arxiv} came to our attention that derives the same law of refraction for ST wave packets in dispersive media in Eq.~\ref{Eq:LoRinDispersive}.

\section*{Funding}
U.S. Office of Naval Research (ONR) N00014-17-1-2458 and N00014-20-1-2789;
Air Force Office of Scientific Research (AFOSR) FA9550-20-1-0322.

\section*{Disclosures}
The authors declare no conflicts of interest.

\bibliography{diffraction}


\end{document}